\documentclass[iop]{emulateapj}
\usepackage{apjfonts}
\newcommand{\be}{\begin{equation}}
\newcommand{\ee}{\end{equation}}
\newcommand{\Msun}{M_{\odot}}
\newcommand\ionn[2]{#1$\;${\scshape{#2}}}
\def\kms{ {\rm km\, s}^{-1}}
\def\pcpc{\ {\rm pc}^{-2}}

\def\Mdyn{M_{\rm dyn}}
\def\Re{R_{\rm e}}

\shortauthors{CONROY, DUTTON, GRAVES, MENDEL, VAN DOKKUM}
\shorttitle{IMF Variation In Compact Early-Type Galaxies}

\begin{document}

%--------------------------------------------------------%
\title{Dynamical Versus Stellar Masses in Compact Early-Type Galaxies:
  Further Evidence for Systematic Variation in the Stellar Initial Mass
  Function}
%--------------------------------------------------------%

\author{Charlie Conroy\altaffilmark{1,2},
  Aaron A. Dutton \altaffilmark{3},
  Genevieve J. Graves\altaffilmark{4}, 
  J. Trevor Mendel\altaffilmark{5}, and
  Pieter G. van Dokkum\altaffilmark{6}}

\altaffiltext{1}{Department of Astronomy \& Astrophysics, University
  of California, Santa Cruz, CA, USA}
\altaffiltext{2}{Alfred P. Sloan Fellow}
\altaffiltext{3}{Max Planck Institute for Astronomy, K\"onigstuhl 17,
  69117, Heidelberg, Germany}
\altaffiltext{4}{Department of Astrophysical Sciences, Princeton
  University, Princeton, NJ, USA}
\altaffiltext{5}{Max-Planck-Institute for Extraterrestrial Physics,
  Giessenbachstrasse, D-85748 Garching, Germany}
\altaffiltext{6}{Department of Astrophysical
  Sciences, Yale University, New Haven, CT, USA}

\slugcomment{Submitted to ApJ Letters}

\begin{abstract}

  Several independent lines of evidence suggest that the stellar
  initial mass function (IMF) in early-type galaxies becomes
  increasingly `bottom-heavy' with increasing galaxy mass and/or
  velocity dispersion, $\sigma$.  Here we consider evidence for IMF
  variation in a sample of relatively compact early-type galaxies
  drawn from the Sloan Digital Sky Survey (SDSS).  These galaxies are
  of sufficiently high stellar density that a dark halo likely makes a
  minor contribution to the total dynamical mass, $\Mdyn$, within one
  effective radius.  We fit our detailed stellar population synthesis
  models to the stacked absorption line spectra of these galaxies in
  bins of $\sigma$ and find evidence from IMF-sensitive spectral
  features for a bottom-heavy IMF at high $\sigma$.  We also apply
  simple `mass-follows-light' dynamical models to the same data and
  find that $\Mdyn$ is significantly higher than what would be
  expected if these galaxies were stellar dominated and had a
  universal Milky Way IMF.  Adopting $\Mdyn\approx M_\ast$ therefore
  implies that the IMF is `heavier' at high $\sigma$.  Most
  importantly, the {\it quantitative amount} of inferred IMF variation
  is very similar between the two techniques, agreeing to within
  $\lesssim0.1$ dex in mass.  The agreement between two independent
  techniques, when applied to the same data, provides compelling
  evidence for systematic variation in the IMF as a function of
  early-type galaxy velocity dispersion.  Any alternative explanations
  must reproduce both the results from dynamical and stellar
  population-based techniques.

\end{abstract}


%--------------------------------------------------------%

\section{Introduction}
\label{s:intro}

The stellar initial mass function (IMF) plays a fundamental role in
studies of distant galaxies.  Stars with masses near the main sequence
turnoff dominate the luminosity of galaxies, and so an IMF is normally
employed to extrapolate the properties of the turnoff stars (e.g.,
their number, age, and metallicity) to the more numerous, but fainter
low mass stars.  With an assumed IMF one can then estimate a variety
of galactic properties, such as the total star formation rate and
stellar mass.  It is standard practice to assume that the IMF is
universal (temporally and spatially invariant) and equal to either the
\citet{Salpeter55} or Milky Way (MW) disk IMF \citep{Kroupa01,
  Chabrier03}\footnote{For practical purposes, in this {\it Letter} we
  equate `the MW IMF' with the \citet{Kroupa01} form.}.  This
assumption is supported by the lack of strong evidence for IMF
variation within the MW \citep{Bastian10}.

This lack of variation has been difficult to reconcile with
theoretical expectations, as nearly every model for the IMF predicts
at least some variation with physical properties, such as the
temperature and pressure of molecular clouds \citep[e.g.,][]{Larson98,
  Larson05, Silk95, Adams96, Padoan97, Padoan02, Hennebelle08,
  Krumholz11, Hopkins12a, Hopkins13a, Narayanan12}.  However, despite
the broad consensus that IMF variation should exist, there is little
agreement on the main variables expected to control the variation, or
even on the qualitative sense of the expected variation.

In the past several years two independent modeling techniques ---
stellar population synthesis (SPS) and dynamical --- have pointed
toward possible variation in the form of the IMF within the
early-type/quiescent galaxy population.  In particular, these
techniques suggest that the IMF becomes increasingly `bottom-heavy'
(i.e., an IMF with a greater proportion of low-mass,
$M\lesssim1\,\Msun$, stars compared to the MW IMF) with increasing
galaxy mass, velocity dispersion, and/or elemental abundance pattern
\citep[e.g.,][]{Cenarro03, vanDokkum10, Treu10, Auger10, Graves10a,
  ThomasJ11, Sonnenfeld12, Cappellari12, Dutton12a, Dutton13a,
  Dutton13b, Conroy12b, Ferreras13, Smith12b, Spiniello12,
  Spiniello13, LaBarbera13, Tortora13}.  At the same time, several
studies have found evidence for `bottom-light' IMFs in low mass
systems \citep{Strader11, Zaritsky13, Geha13}, and perhaps even in the
Small Magellanic Cloud \citep{Kalirai13}.  In the most extreme cases
the quoted variation in the IMF amongst massive early-type galaxies
would imply a revision in the stellar masses upward by a factor of
$3-4$.

Despite the broad qualitative agreement between two independent
techniques, there is still some room for skepticism.  Both dynamical
and SPS techniques are subject to a variety of systematic
uncertainties that are difficult to quantify.  Only very recently have
several groups begun to make direct comparisons between dynamical and
SPS constraints on IMF variation \citep{LaBarbera13, Tortora13}.  The
goal of this {\it Letter} is to directly compare dynamical and
SPS-based mass-to-light ratios for a sample of compact early-type
galaxies.  We expect these galaxies to be stellar-dominated, at least
within an effective radius.  Consideration of compact galaxies is
therefore expected to reduce an important source of systematic
uncertainty in the dynamical models (i.e., the inclusion of a dark
halo).

The rest of this {\it Letter} is organized as follows.  In Section
\ref{s:data} we describe the data and sample selection; in Section
\ref{s:methods} we provide a brief overview of the dynamical and SPS
techniques.  The results are presented in Section \ref{s:res}.  Where
necessary a flat $\Lambda$CDM cosmology is assumed with the following
parameters: $(\Omega_{\rm m},\Omega_\Lambda,H_0) = (0.3,0.7,70)$.

%--------------------------------------------------------%

\begin{figure}[!t]
\center
\resizebox{3.5in}{!}{\includegraphics{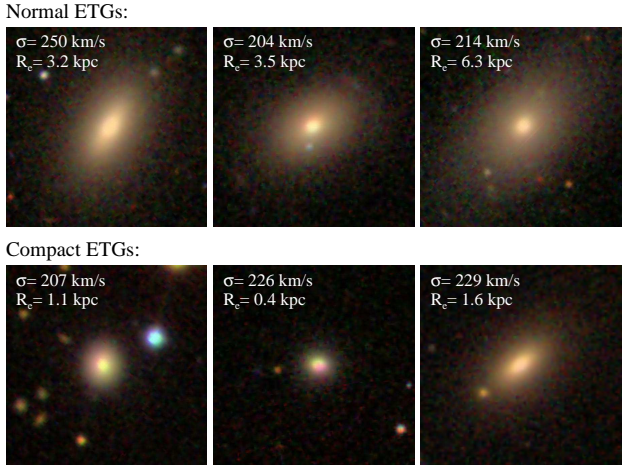}}
\caption{Representative $gri$ composite images of early-type galaxies
  in a narrow $\sigma$ bin.  Each image is
  $60\arcsec\times60\arcsec$. These galaxies were chosen to reside in
  a narrow redshift interval, $0.03<z<0.04$, so that the relative
  apparent sizes are equal to relative physical sizes.  The upper and
  lower panels compare normal and compact early-type galaxies (ETGs).}
\label{fig:img}
\end{figure}

\section{Data \& Sample Selection}
\label{s:data}

The sample of compact early-type galaxies analyzed in this {\it
  Letter} is drawn from the SDSS \citep{York00} Data Release 7
\citep{abazajian09}.  Stellar velocity dispersions, $\sigma$, are
taken to be a weighted mean of the ``SDSS'' and ``Princeton''
spectroscopic pipelines.  Total circularized half-light radii, $\Re$,
are measured via bulge plus disk decompositions provided in
\citet{Simard11}.  As a check on these half-light radii, we also
consider half-light sizes provided by the SDSS photometric pipeline
(specifically, the sizes based on a de Vaucouleurs light profile).
All sizes are corrected for seeing.  Photometric stellar mass-to-light
ratios, $M_{\ast,\rm MW}/L$, are available for this sample from two
sources: a catalog produced by one of us (Mendel et al. in prep), and
a catalog provided by the MPA/JHU group (available at
http://www.mpa-garching.mpg.de/SDSS/DR7/).  The Mendel et al. stellar
masses were derived by modeling the SDSS $ugriz$ total magnitudes
(based on the Simard et al. light profile fits) employing standard
population synthesis techniques.  In both catalogs, stellar
mass-to-light ratios were also computed from the fiber-based
photometry in order to compare directly with spectroscopically-based
values.  In the photometric analyses a \citet{Chabrier03} IMF was
adopted.  We have converted these masses to a \citet{Kroupa01} IMF by
increasing them by 10\%, the approximate conversion between the two
IMFs for old stellar populations.  The quantity $M_{\ast,\rm MW}$ will
throughout refer to stellar masses assuming a MW \citep{Kroupa01} IMF.

We focus on compact galaxies in this work because there are several
reasons to believe that they should be stellar-dominated at least
within an effective radius.  First, their densities are so high that
one would require substantial modification to a standard NFW
\citep{NFW97} dark matter density profile in order to achieve central
dark matter densities high enough to rival the stellar densities.  If
these compact galaxies resided within NFW dark matter halos of mass
$M_h=10^{12.5}\,\Msun$ with a concentration of $c=6$ \citep[expected
from the $c(M)$ relation;][]{Maccio08}, the dark matter fraction,
$f_{\rm DM}$, within $R_e$ would be 5\%.  Considering concentrations
that lie $2\sigma$ above the concentration-mass relation ($c=10$)
yields $f_{\rm DM}=10$\%.  Increasing the halo mass to
$10^{13}\,\Msun$ changes these fractions to 7\% and 13\%.  These
numbers do not include the possibility of adiabatic contraction.  We
will show in Section \ref{s:res} that a dark matter fraction of 65\%
would be required to explain the totality of the trend reported here.
This is incompatible with kinematic models for these galaxies.  They
follow the virial fundamental plane, and their observed scaling
relations and two-dimensional kinematics can only be fit with
mass-follows light models, with an upper limit on the central DM
fractions of <20\% \citep[see][for details]{Dutton12a, Dutton13a,
  Cappellari13a}. When modeling $\Mdyn$ we make the simplifying
assumption that these galaxies are stellar dominated and hence that
``mass follows light''.

\begin{figure}[!t]
\center
\resizebox{4in}{!}{\includegraphics{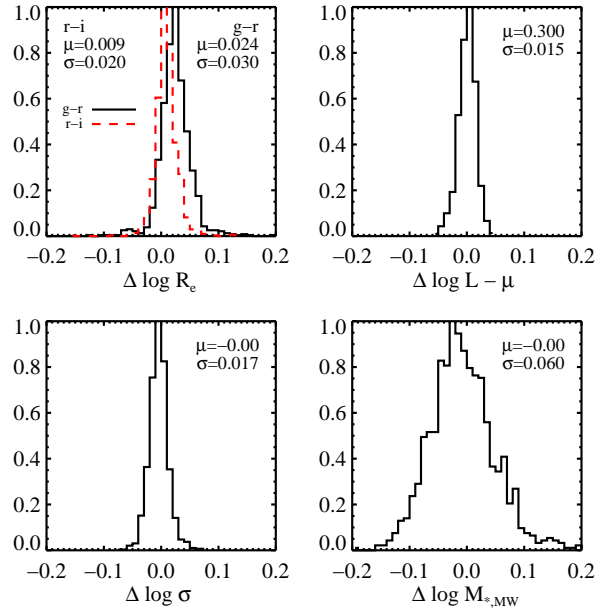}}
\caption{Stability of the physical properties of the sample.  In each
  panel, the mean and standard deviation of each distribution is shown
  in the legend.  {\it Top Left:} Difference in $\Re$ between the $g$,
  $r$, and $i$ bands, as determined by the SDSS photometric pipeline
  assuming a de Vaucouleurs light profile.  {\it Top Right:}
  Difference in total luminosity measured in the $g$ and $r$
  bands. {\it Bottom Left:} Difference in velocity dispersion measured
  from the same spectra but with two different reduction pipelines
  \citep[the ``SDSS'' and ``Princeton'' versions; see also][]{Hyde09}.
  {\it Bottom Right:} Difference in stellar masses derived from
  photometry and assuming a \citet{Kroupa01} IMF, for two different
  fitting codes (Mendel et al. and MPA/JHU).}
\label{fig:prop}
\end{figure}

\begin{deluxetable}{ccccccc}
\tablecaption{Properties of Compact Early-Type Galaxies}
\tablehead{ \colhead{log $\sigma$} &\colhead{N} & 
\colhead{$R_{\rm fib}/\Re$} &\colhead{$\Re$} &  
\colhead{$M_{\ast,{\rm MW}}$}  & \colhead{$\Mdyn/M_{\ast,{\rm MW}}$ } &
\colhead{$M_\ast/M_{\ast,{\rm MW}}$} }
\startdata
1.96 &  34 &  1.9 &  0.6 & 10.0 &  0.98$\pm$  0.05&  1.28$\pm$  0.18\\
2.06 & 172 &  1.7 &  0.8 & 10.2 &  1.24$\pm$  0.02&  1.35$\pm$  0.09\\
2.15 & 352 &  1.5 &  0.9 & 10.4 &  1.55$\pm$  0.02&  1.47$\pm$  0.07\\
2.25 & 345 &  1.2 &  1.2 & 10.6 &  1.94$\pm$  0.03&  1.69$\pm$  0.06\\
2.34 & 169 &  1.0 &  1.4 & 10.7 &  2.31$\pm$  0.05&  1.85$\pm$  0.08\\
2.43 &  32 &  0.7 &  1.9 & 11.0 &  2.90$\pm$  0.15&  2.27$\pm$  0.16
\enddata
\vspace{0.1cm} 
\tablecomments{All quantities are averages within the bin except for
  $N$, the number of galaxies in each bin.  Velocity dispersion,
  $\sigma$, is in units of log $\kms$, $R_{\rm fib}/\Re$ is the
  average size of the fiber in units of the half-light radius,
  $r-$band half-light radius, $\Re$, is in kpc, and $M_{\ast,{\rm MW}}$
  is quoted in units of log $\Msun$.  Errors are $1\sigma$
  statistical uncertainties.}
\label{t:data}
\end{deluxetable}

Our sample of compact early-type galaxies was selected according to
the following cuts.  A redshift range of $0.025<z<0.06$ was imposed so
that the reddest part of the CaT spectral feature is always in the
observed window.  Galaxies were selected to have a spectroscopic
pipeline flag eCLASS$<0$, indicating an early-type spectrum, a red
$g-r$ color ($g-r>0.59+0.052\,{\rm log} [M_{\ast,\rm{MW}}/\Msun-10]$),
and a minor-to-major axis ratio greater than 0.5.  These cuts result
in a sample of 19,000 galaxies.  We then computed stellar surface mass
densities within $R_e$ according to: $\Sigma_\ast=M^{\rm
  \,tot}_{\ast,\rm MW}/(2\pi\,\Re^2)$, where $\Re$ is measured in the
$r-$band and $M^{\rm \,tot}_{\ast,\rm MW}$ is the
photometrically-derived total stellar mass assuming a MW IMF.
Galaxies were selected to be compact via the cut
$\Sigma_\ast>2500\,\Msun\pcpc$.  These cuts yield 1100 compact
early-type galaxies ($\approx6\%$ of the early-type galaxy population)
that define the sample used in the present analysis.  In the following
analysis these galaxies are grouped into six equally-spaced bins in
log$\,\sigma$ from $1.9<{\rm log}\,\sigma/\kms<2.5$.  Several derived
parameters in these $\sigma$ bins are shown in Table 1.  Figure
\ref{fig:img} shows $gri$ composite images for both normal and compact
early-type galaxies in a narrow bin in redshift and $\sigma$.  On
average, the compact galaxies are $2.5\times$ smaller than the overall
early-type population, implying that they are $\approx15\times$ denser
(in terms of $\Msun\,{\rm pc}^{-3}$).

These galaxies are remarkably dense and small; from the lowest to the
highest $\sigma$ bins, the mean sizes range from 0.6 kpc to 2.3 kpc,
corresponding approximately to $0.8\arcsec-2.3\arcsec$ at the median
redshift of the sample (see Table 1).  For reference, the SDSS pixel
scale is $0.4\arcsec$ and the seeing is $1\arcsec-1.5\arcsec$.  Only
the largest galaxies in our sample are therefore (just) resolved
within the half-light radius.  Note however that half-light
sizes can be reliably estimated even for such small galaxies because
their total sizes subtend a considerably larger angle on the sky (see
Figure \ref{fig:img}).

The SDSS obtains spectra with fibers that have $3\arcsec$ diameters.
Despite the fact that our sample is of relatively low redshift, their
high surface densities imply that the SDSS fibers cover a significant
fraction of the galaxy.  From the lowest to the highest $\sigma$ bins
the fiber covers on average the inner 1.9$\,\Re$ to 0.7$\,\Re$ (see
Table 1).  When fitting the data to SPS models we utilize spectra that
are stacked in bins of $\sigma$.  Each spectrum is
continuum-normalized and broadened to a common dispersion of
$\sigma=350\,\kms$, and the stack is created by giving equal weight to
each spectrum, after masking strong sky lines. The resulting stacked
spectra have signal-to-noise ratios ranging from $\sim100-700$
\AA$^{-1}$.

In Figure \ref{fig:prop} we show the robustness of the parameters used
to select the compact early-type galaxy sample. The key point is that
sizes, luminosities, velocity dispersions, and stellar masses (with an
assumed MW IMF) are all measured consistently between different
images, in different bands, and/or using different photometric
pipelines. Therefore, even though we have selected outliers (i.e.,
extremely com- pact galaxies), their properties are robust and not due
to measurement errors.  Note that the derivations of $\Mdyn/L$ shown
below use the full galaxy light profiles, and thus do not depend
explicitly on any of these measured parameters

%----------------------------------------------------------%

\begin{figure}[!t]
\center
\resizebox{3.6in}{!}{\includegraphics{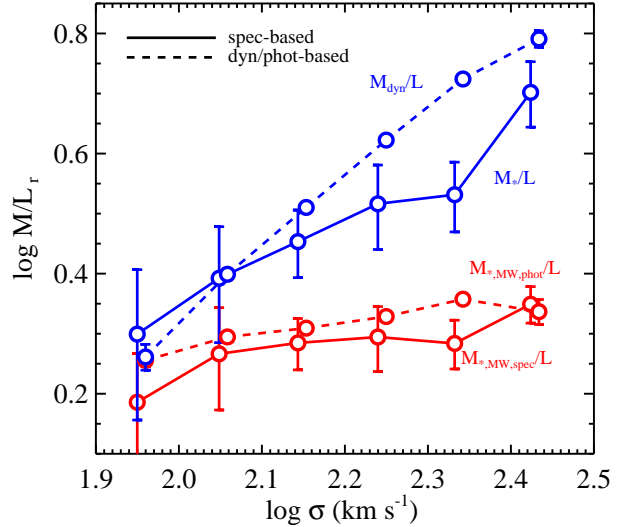}}
\caption{Mass-to-light ratio in the $r$-band, $M/L_r$, as a function
  of galaxy velocity dispersion, $\sigma$.  Values obtained from
  fitting absorption line spectra (solid lines) are compared to values
  estimated from photometric SPS and dynamical modeling (dashed
  lines).  Red lines are results for a fixed MW IMF, while blue lines
  make no assumption about the IMF.  In particular, $M_\ast/L$ is the
  result of allowing the IMF to be constrained directly by the
  absorption line data.  The difference between $M_\ast/L$ and
  $M_{\ast, {\rm MW, spec}}/L$ is due solely to differences in the
  IMF, while the difference between $\Mdyn/L$ and $M_{\ast, {\rm MW,
      phot}}/L$ can in principle be due to either the IMF or dark
  matter.  Errors on the dashed lines are very small because they are
  errors on the mean, whereas the errors on the solid lines reflect
  the uncertainty from fitting stacked spectra.}
\label{fig:mli}
\end{figure}

\begin{figure*}[!t]
\center
\resizebox{6in}{!}{\includegraphics{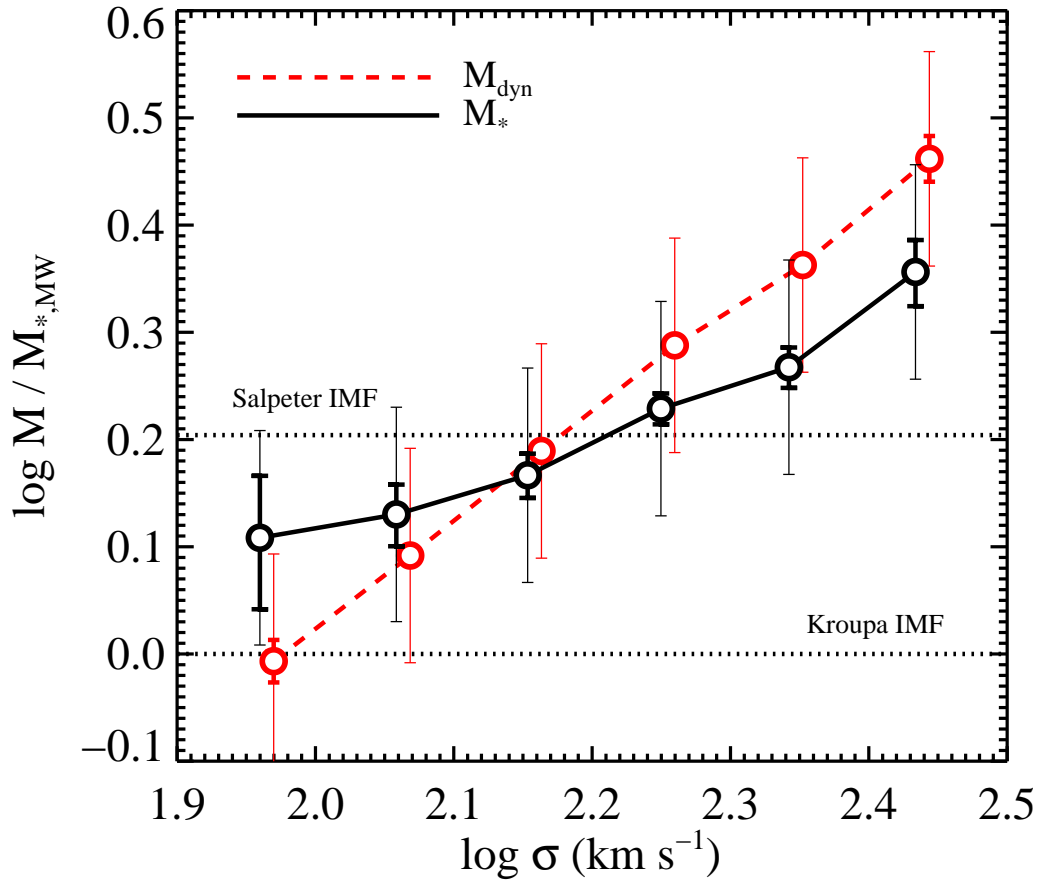}}
\caption{Variation in the normalized mass with galaxy velocity
  dispersion, $\sigma$, where `normalized mass' is defined as the
  measured mass in units of the stellar mass derived assuming a MW
  \citep{Kroupa01} IMF.  Dynamical masses (dashed line) are compared
  to stellar population-based masses (solid line).  For the latter,
  the $y-$axis is sensitive {\it only} to the IMF, while for the
  former the $y-$axis is in principle sensitive to both IMF variation
  and dark matter.  Dotted lines show the expected values if the
  galaxies had \citet{Kroupa01} or \citet{Salpeter55} IMFs.  Thick
  error bars represent formal statistical errors, while the thin error
  bars represent 0.1 dex systematic errors.  The two curves are
  shifted slightly along the $x-$axis for clarity.}
\label{fig:res}
\end{figure*}

\vspace{1cm}

\section{Methods}
\label{s:methods}

In this section we describe the techniques for computing $\Mdyn/L$ and
$M_\ast/L$ for our sample of compact early-type galaxies.

Stellar mass-to-light ratios are computed by fitting an SPS model to
the stacked absorption line spectra of the compact early-type
galaxies.  The SPS model used herein was developed in
\citet[][]{Conroy12a,Conroy12b}.  In its present form the model
contains 27 free parameters, including the redshift and velocity
dispersion, a two-part power-law IMF, two population ages, four
nuisance parameters, and the abundances of 17 elements.  These
parameters are fit to the data via a Markov Chain Monte Carlo fitting
technique.  For the purposes of this {\it Letter} the most important
feature is that the IMF is constrained directly by the data, in
particular by a variety of spectral features that lie mainly in the
red spectral range (e.g., numerous TiO bands at $>6000$\AA,
\ionn{Na}{i} at 8200\AA\, and \ionn{Ca}{ii} at $8500-8670$\AA).  The
data and models are split into four wavelength intervals and within
each interval the spectra are normalized by a high-order polynomial
\citep[see][for details]{Conroy12b, Conroy13a}.  The main result of
this fitting procedure is an estimate of the true stellar
mass-to-light ratio, $M_\ast/L$; `true' because the IMF is included as
a parameter determined by the data.  For each set of model parameters
we also keep track of the stellar mass-to-light ratio that would be
expected assuming a MW \citep{Kroupa01} IMF, $M_{\ast,\rm MW}/L$.
Notice that the ratio between the measured stellar mass and the
stellar mass assuming a MW IMF, $M_\ast/M_{\ast,\rm MW}$, is sensitive
only to the IMF.  The resulting best-fit parameters are very
well-constrained, with essentially no degeneracies between the IMF and
other parameters.  Quoted errors are $1\sigma$ limits marginalized
over the full posterior distribution.

For each galaxy we calculate $\Mdyn$ and $\Mdyn/L$ from the following
procedure \citep[see][for details]{Dutton12a}. We start with a galaxy
light profile parameterized by circularized and de-projected S\'ersic
$n=4$ and $n=1$ components.  Then, assuming that mass follows light,
we calculate the radial velocity dispersion profile by solving the
spherical Jeans equations assuming isotropic orbits \citep[$\beta=0$;
e.g.,][]{Binney82} and a purely pressure-supported system.  As
discussed in \citet{Dutton12a}, due to the large fraction of the
galaxy light covered by the fiber, the dynamical masses are only
weakly dependent on stellar anisotropy \citep[see also][]{Wolf10}.
Adopting $\beta=0.5$ changes $\Mdyn$ by $<5$\%.  We then compute the
projected velocity dispersion (convolved with $1.4\arcsec$ FWHM
seeing, typical for the SDSS) within the $3\arcsec$ diameter aperture
used by SDSS. Finally, we compute $\Mdyn$ by scaling the model
aperture velocity dispersion to match the observed aperture velocity
dispersion.  Notice that the dynamical masses are completely
independent of the SPS stellar masses, and since mass follows light,
$\Mdyn/L$ is constant with radius.  We then compute the mean $\Mdyn/L$
within the SDSS fiber aperture in bins of $\sigma$.  In addition to
the formal statistical error, we also assign a systematic error of 0.1
dex to the dynamical masses.  The motivation for this error comes in
part from Figure \ref{fig:prop}, where sizes and dispersion can differ
by $0.02-0.06$ dex depending on assumptions.  In addition,
uncertainties in the anisotropy profile, degree of rotational support,
etc., when combined can be expected to affect $\Mdyn/L$ at the 0.1 dex
level.

There are four distinct `mass-to-light ratios' in this {\it Letter},
which we summarize here for clarity.  $\Mdyn/L$ is based on a
dynamical model; $M_\ast/L$ is reserved for the `true' stellar
mass-to-light ratio inferred from spectral fitting when the IMF is
left as a free parameter; $M_{\ast, {\rm MW}}/L$ is, generically, the
stellar mass-to-light ratio assuming a MW IMF.  In the present work
there are two estimates of this last quantity; the first from fitting
$ugriz$ photometry, and the second from fitting the absorption line
spectra with the IMF fixed to the MW form.  In practice these two
techniques provide nearly identical estimates of $M_{\ast, {\rm MW}}/L$,
so we do not distinguish between these two except where necessary.

Finally, notice that all quoted mass-to-light ratios are determined
within the same aperture, specifically within the $3\arcsec$ SDSS
fiber.  `Total' mass-to-light ratios and masses are only used for
selecting the sample.

%---------------------------------------------------------%

\vspace{1cm}

\section{Results}
\label{s:res}

The best-fit mass-to-light ratios, $M/L$, for the three techniques ---
photometric, spectroscopic, and dynamical --- are shown in Figure
\ref{fig:mli}.  The red lines compare photometrically and
spectroscopically derived $M_{\ast, {\rm MW}}/L$ values, and it is
encouraging that they agree well.  The blue lines compare the results
from dynamical models to the results from the spectroscopic model,
where in the latter case the IMF is a free variable derived from the
data.  All $M/L$ ratios in this figure were derived within the same
aperture (the spectroscopic fiber).  As we argued in Section
\ref{s:data}, the expected contribution from dark matter should be
small in these compact galaxies, certainly less than the dark matter
fractions of $\approx65$\% that would be required to attribute
entirely to dark matter the difference between $\Mdyn/L$ and $M_{\ast,
  {\rm MW}}/L$ at high $\sigma$.

The main result of this {\it Letter} is shown in Figure \ref{fig:res}.
In this figure we compare the best-fit dynamical and SPS-based masses
as a function of galaxy velocity dispersion, $\sigma$.  The key
feature of this diagram is that masses are plotted in units of the
stellar mass assuming a MW IMF, $M_{\ast,\rm MW}$.  The $y$-axis is
thus insensitive to normal stellar population variation such as
stellar age, metallicity, etc.  Dotted lines show the expected ratios
if the galaxies had \citet{Kroupa01} or \citet{Salpeter55} IMFs.  In
general $\Mdyn$ will be a mix of stellar mass and dark halo mass, but
as we argued in Section \ref{s:data}, the galaxies analyzed herein are
sufficiently dense that $\Mdyn$ is likely dominated by the stars.  We
also emphasize that $M_\ast/L$ and $\Mdyn/L$ are completely
independent in the sense that the former is derived from absorption
line spectra while the latter is derived from a combination of $\Re$,
$L$, $\sigma$, and the detailed light profile shape.  Furthermore,
while formally the axes are correlated for the $\Mdyn/M_{\ast,\rm MW}$
vs. $\sigma$ relation ($\Mdyn\propto\sigma^2$), we have simulated the
effect of measurement errors and find that the nominal measurement
errors (Figure \ref{fig:prop}) have a small effect on the result.  If
the true random errors were as large as 0.05 dex, then half of the
difference between the $\Mdyn$ and $M_\ast$ relations could be
explained by the correlated axes in the former relation.

The key result is that $\Mdyn/M_{\ast,\rm MW}$ and $M_\ast/M_{\ast,\rm
  MW}$, and their variation with $\sigma$, agree very well, perhaps
even better than expected given expected systematic errors that could
plausibly exceed 0.1 dex.  In addition, both masses point toward an
IMF that becomes increasingly bottom-heavy with increasing $\sigma$.
At the highest dispersions the inferred IMF becomes considerably
steeper than even the Salpeter IMF.

We emphasize that not only were the dynamical, photometric and
spectroscopic masses derived for the same set of galaxies, but they
were also derived for data obtained within the exact same aperture,
i.e., the (seeing-convolved) SDSS fiber.  Figure \ref{fig:res}
therefore presents the most direct comparison of spectral and
dynamically-based IMF results to-date.

We caution that the quantitative trend shown here for our sample of
1100 compact early-type galaxies may not to be representative of the
entire early-type galaxy population, and in fact there are emerging
clues that galaxy compactness may be an important variable governing
the IMF shape \citep{Lasker13, Smith13}.  In a forthcoming analysis we
will explore the two dimensional manifold of IMF variation as a
function of $\sigma$ and galaxy compactness (e.g., $\Sigma_\ast$
and/or $\Re$).

The fact that the same IMF variation is seen both via dynamical and
SPS-based techniques is reassuring and suggests that neither technique
suffers from catastrophic systematic errors.  However, we caution that
consistency does not necessarily imply that both techniques are
correct, as independent systematic errors in the mass measurements
could work in the same direction.  Addressing possible sources of
systematic uncertainty is the subject of ongoing work.

%---------------------------------------------------------%

\acknowledgments 

CC acknowledges support from the Alfred P. Sloan Foundation.   This
paper is based on data from the Sloan Digital Sky Survey.

\end{document}